\newcommand{\be}{\begin{equation}}
\newcommand{\e}{\end{equation}}
\newcommand{\f}{\frac}
\newcommand{\de}{{\rm d}}
\newcommand{\hmpc}{$h^{-1}$\,Mpc}

\documentstyle[psfig,astrobib,emulateapj]{article}

\begin{document}
\submitted{Accepted in ApJ}
\title{A simple analytical model for the abundance of 
damped Ly$\alpha$ absorbers}
\author       {T. Roy Choudhury\thanks{E-mail: tirth@iucaa.ernet.in},
	T. Padmanabhan\thanks{E-mail: paddy@iucaa.ernet.in}}
\affil{IUCAA, Post Bag 4, Ganeshkhind, Pune 411 007, India.}

\begin{abstract}
A simple analytical model for estimating the fraction ($\Omega_{\rm gas}$) 
of matter  
in gaseous form within the collapsed dark matter (DM) haloes is presented. 
The  model is developed using  (i) the Press-Schechter formalism 
to estimate the fraction of baryons in DM haloes, 
and (ii) the observational estimates 
of the star formation rate at different redshifts. The prediction for 
$\Omega_{\rm gas}$ from the model is in broad agreement with 
the observed abundance of the damped Ly$\alpha$ systems. 
Furthermore, it can be used for estimating the 
circular velocities of the collapsed haloes at different redshifts, which 
could be compared with future observations. 
\end{abstract}

\keywords{ 
 - cosmology: large-scale structure of universe
-- intergalactic medium -- quasars: absorption lines
}

\section{Introduction}

The Damped Lyman-$\alpha$ systems (DLA) are identified with 
the lines having highest 
column densities in a typical observed absorption spectrum of a 
distant quasar. The generally adopted threshold value of the 
column density ($N_{\rm HI}$) for identifying DLAs is 
$N_{\rm HI} \ge 2 \times 10^{20}$ cm$^{-2}$. These high 
column density systems are important in understanding 
the baryonic structure formation because they 
contain fair amount of the neutral hydrogen in the universe at 
high redshifts \cite{wtsc86,smi96}. In contrast,  
the ionised hydrogen is mostly contained 
in the low column density systems, which manifest themselves 
as the Ly$\alpha$ forest 
in the quasar absorption spectrum
(see, for example, \citeNP{mcor96,bd97,cps01,csp01} and references therein).
By probing the DLAs at high redshifts, one is able 
to extract information about formation and evolution 
of galaxies. In recent years, the DLAs have been studied extensively 
using high quality 
quasar absorption spectra 
\cite{pshk94,lwt95,smi96,pw97,pksh97,pskh97,pw98,pesb99}, 
which has helped in constraining 
galaxy formation models.

Currently, the theoretical understanding of the DLAs is based on 
the hierarchical models for galaxy formation
\cite{wr78,wf91}. 
These models start with the fact that, initially, 
the DM density 
inhomogeneities collapse via gravitational instability 
and form potential wells. In the next stage
the baryonic matter, mostly in gaseous form, 
follow these DM potential wells. 
The baryonic gas can cool and form galaxies, provided its 
cooling timescale is short compared to the Hubble expansion timescale. 
Once this 
condition is met, the gas is able to 
dissipate its energy and `virialise' in the centres of the DM haloes.
This cooled gas, contained within the DM haloes, is believed to be 
responsible for the damped Ly$\alpha$ absorption lines in the quasar 
spectra. Various analytical and semi analytical models 
\cite{mm94,kc94,sp94,jp98,kauffmann96,mpsp01}
as well as simulations \cite{mb94,kwhm96,mbhwk97,gkhw97,gkwh97,hsr98}, 
have shown that the  
observed mass in the DLAs is similar to the baryonic mass which can collapse 
and efficiently cool in the DM haloes. However, not all the collapsed 
baryonic matter  
in the haloes remain in gaseous form -- a fraction will turn into luminous 
stars. It is, therefore, clear that estimation of the total 
baryonic mass in the collapsed haloes and the mass turned into stars 
can be used to calculate the mass of the cooled gas in the haloes.

In this paper, we use the simple idea that the mass contained in DLAs 
(i.e., the mass which remains in gaseous form in the collapsed haloes) 
should be equal to the total baryonic mass in the collapsed haloes minus 
the mass which is turned into stars. In doing this, we have neglected 
the mass of the ionised gas within the haloes (which is small 
because the gas is dense enough to be shielded from the background 
radiation) and also the small amount of mass which might have 
turned into molecular gas.
The total baryonic mass contained within the collapsed haloes can be 
estimated through Press-Schechter formalism \cite{mm94,kauffmann96}, 
while the 
matter which has turned into stars is calculated from the observed 
estimates of the star formation rate (SFR) at various epochs \cite{sagdp99}. 
It turns out that this simple model is able to produce the mass of the 
cooled gas in the collapsed haloes as seen in observations. Further, 
it is able to predict the circular velocities of the 
collapsed haloes associated 
with DLAs. 

This simple model should not be thought of as a competitor to the 
previous detailed models which use a lot more inputs for their modelling. 
Rather, we have tried to show that it is possible to reproduce many 
of the physical
effects through a simple argument of `baryon conservation'.
Our paper should be judged in this backdrop of simplicity.

The paper is arranged as follows. The next section discusses the basic 
analytical formalism used for the model. It describes the usage 
of the Press-Schechter 
formalism to calculate the total baryonic mass in collapsed haloes. It 
also discusses how to obtain the stellar mass density from the 
observed SFR. In section \ref{sec:results} we discuss the parameters 
of our model, followed by the results. In the last section, we summarise 
our main conclusions, and discuss the limitations of our model.

\section{Analytical Formalism}
\label{sec:analytic}

This section contains the basic formalism for the model. Although most of it 
is available in existing literature, we briefly summarise them for  
setting up the notation and for stating the assumptions used 
in our model. To calculate the mass contained within the collapsed objects 
of a given mass range, 
we use the Press Schechter formalism
\cite{ps74}. 
The total mass (dark matter) contained per unit comoving volume 
within collapsed objects of (logarithmic) mass range 
$(\ln M, \ln(M+\de M))$ is given by
\begin{eqnarray}
\rho_M^{\rm halo}(z) ~\de \ln M &=&
-\left(\sqrt{\f{\pi}{2}} {\rm e}^{-\nu^2/2}\right)
\rho_0 
\f{\de \ln \sigma(M)}{\de \ln M} \nonumber\\
&\times& \nu(z,M) ~\de \ln M
\end{eqnarray}
where $\rho_0$ is the mean comoving density of the 
universe. The other quantities are defined as
\begin{eqnarray}
\sigma^2(R) &=& \int_0^{\infty} \f{\de k}{k} \left[\f{3 (\sin kR-kR
\cos kR)}{k^3 R^3}\right]^2 
\left[\f{k^3 P_{\rm DM}(k)}{2 \pi^2}\right]; 
\nonumber \\
M&=&\f{4\pi}{3} R^3 \rho_0
\end{eqnarray}
The quantity $\nu$ is defined as
\be
\nu(z,R)=\f{\delta_c}{D(z) \sigma(R)}.
\e
$P_{\rm DM}(k)$ is the normalised dark matter power spectrum, $D(z)$ is 
the growth factor for density perturbations and 
$\delta_c$ is the critical density, usually 1.69 for $\Omega_m=1$ 
flat universe. 
The corresponding $\Omega$ is given by
\be
\Omega_M^{\rm halo}(z)\equiv \f{\rho_M^{\rm halo}(z)}{\rho_c} 
= \f{\rho_M^{\rm halo}(z) \Omega_m} {\rho_0}
\e
where $\rho_c=2.8 \times 10^{11} h^2 ~ M_{\odot}$ Mpc$^{-3}$ is 
the present critical density of the universe and  
$\Omega_m=\rho_0/\rho_c$.

We shall now calculate the total mass contributed by {\it baryons} 
in the collapsed haloes within the logarithmic mass range. In order 
to do this, 
we assume that the baryonic fraction of matter in each halo is same 
as the global value. Then 
the $\Omega$ contributed by baryons in collapsed haloes within 
a logarithmic mass range $(\ln M, \ln(M+\de M))$ is given by
\begin{eqnarray}
\Omega_{M,b}^{\rm halo}(z) &=& 
\Omega_M^{\rm halo}(z) \f{\Omega_b}{\Omega_m} \nonumber\\
&=& -\left(\sqrt{\f{\pi}{2}} {\rm e}^{-\nu^2/2}\right) \nu(z,M)
\f{\de \ln \sigma(M)}{\de \ln M} \Omega_b
\label{omega_m_b_z}
\end{eqnarray}

We now have the total baryonic $\Omega_{M,b}^{\rm halo}$ 
contained within collapsed dark matter 
haloes of mass $M$ at a given epoch. We will assume 
that all the baryons are either converted into luminous stars, or 
they form gaseous clouds. Mathematically, this can be expressed as
\be
\Omega_{M,b}^{\rm halo}(z) = \Omega_{M,{\rm gas}}(z) + \Omega_*(z)
\label{omega_m_b_halo_z}
\e
It is believed that the damped Ly$\alpha$ systems are mainly contributed 
by the gaseous clouds {\it within the collapsed and virialised} 
dark matter 
haloes. Hence, one should {\it not} include the low column 
density systems in this analysis, which are believed to be 
density perturbations 
within a diffuse intergalactic medium (IGM) \cite{cps01,csp01}.

It is clear that once we obtain the quantity $\Omega_*$ contained 
in stars, we can estimate $\Omega_{M,{\rm gas}}$ from the above relation.
To calculate $\Omega_*$ from observational data, we proceed as follows:
The star formation rate (SFR) 
$(\de \rho_*/\de t) \equiv \dot{\rho_*}(t)$ 
is defined as the rate 
at which baryonic mass is converted into stars per unit (comoving) 
volume. Given this quantity, we can obtain the total density of matter 
contained within stars at a particular $t$
\be
\rho_*(t)=\int_0^t \de t~\dot{\rho_*}(t)
\e
or, in terms of redshift
\be
\rho_*(z)=\int_{\infty}^z \de z~\dot{\rho_*}(z) \f{\de t}{\de z}
         =\int_z^{\infty} \de z~\f{\dot{\rho_*}(z)}{H(z) (1+z)}
\label{rho_star_rhodot}
\e
The corresponding $\Omega$ is
\be
\Omega_*(z)=\f{\rho_*(z)}{\rho_c}
\label{omega_star_rho_star}
\e
One can determine $\dot{\rho_*}(t)$ from observations, 
which can then be integrated to give $\Omega_*(z)$. It is well known that 
not all the gas which is turned
into stars is removed from the gaseous phase forever. Actually, 
some of the stellar material is returned by stellar winds and supernovae.
However, we shall ignore this contribution for the following reasons: 
(i) It is difficult to implement this feedback effect in our model 
without introducing more free parameters \cite{efstathiou00}. 
This spoils the simplicity as well 
as the predictive capacity of our model. (ii) The SFR has 
large uncertainties at high redshifts due to extinction 
(see below). 
The correction due the feedback through stellar winds and supernovae 
is less than this uncertainty in the SFR, and can be ignored as a 
first approximation.

To compare our results with observations, or to discuss any observational 
consequences of our model, it is better to work in terms of the circular 
velocities ($v_c$) 
of the collapsed haloes, rather than in terms of their masses. 
The mass $M$ and $v_c$ can be related to each other using spherical 
collapse model. The relevant equations for a universe with 
cosmological constant are \cite{sp99}
\be
v_c^2=\f{GM}{r_{\rm vir}}-
\f{\Omega_{\Lambda} ~ H_0^2 ~ r_{\rm vir}^2}{3}
\e
and
\be
\Delta_{\rm vir}(z)=\f{2GM}{r_{\rm vir}^3 H^2(z)}
\e
where $\Delta_{\rm vir}(z)$ is the virial density of the collapsed halo 
at redshift $z$ and $r_{\rm vir}$ is the corresponding virial radius. The 
circular velocity is calculated assuming that the virialised halo 
has a singular 
isothermal density profile $\rho(r) \propto r^{-2}$.
The 
above relations can be used to eliminate $r_{\rm vir}$ and obtain
\begin{eqnarray}
\f{M}{10^{11} ~ h^{-1} M_{\odot}} &=& 
\left(\f{v_c}{35.0 ~ {\rm km~s^{-1}}}\right)^3 ~
\sqrt{\f{2 H_0^2}{H^2(z) \Delta_{\rm vir}(z)}} \times \nonumber \\
& \times & 
\left(1-
\f{2 \Omega_{\Lambda} H_0^2}
{3 H^2(z) \Delta_{\rm vir}(z)} \right)^{-3}
\label{mass_vel}
\end{eqnarray}
Thus, for a given $z$ we can relate $M$ to $v_c$, provided we know 
$\Delta_{\rm vir}(z)$. This result 
can then be used for obtaining the mass density, 
either dark matter ($\Omega_{v_c}^{\rm halo}(z)$) or baryonic 
($\Omega_{v_c,b}^{\rm halo}(z)$), contributed 
by haloes having a circular velocity $v_c$.

Given the 
background cosmological model, our formalism can now be used 
for estimating the quantity 
$\Omega_{v_c,{\rm gas}}(z)$ for different values of $v_c$, which can then 
be compared with observations. 
{\it Note that we have only one free parameter 
in our model}, namely $v_c$. 
It turns out that we can, in principle, predict 
the circular velocities of collapsed haloes at different redshifts 
by comparing our model with DLA observations.

\section{Model Parameters and Results}
\label{sec:results}

We have considered two different cosmological models with the 
parameters listed below: 

SCDM: $\Omega_m=1, \Omega_{\Lambda}=0, h=0.65, \Omega_b h^2=0.02$

LCDM: $\Omega_m=0.3, \Omega_{\Lambda}=0.7, h=0.65, \Omega_b h^2=0.02$

\noindent 
The critical density for collapse $\delta_c$ is only weakly 
dependent on the $\Omega_{\Lambda}$ for flat cosmological models
\cite{ecf96}; hence 
we take it to be 1.69 for both the models.
The next cosmological input that is required is the form of the DM power 
spectrum. We  take the following form for $P_{{\rm DM}}(k)$ 
\cite{ebw92}
\be
P_{{\rm DM}}(k)=\f{A k}{(1 + [\alpha k + (\beta k)^{1.5} + 
(\gamma k)^{2}]^{\nu})^{2/\nu}}
\e
where $\nu=1.13$, 
$\alpha=(6.4/\Gamma)$\hmpc, $\beta=(3.0/\Gamma)$\hmpc, 
$\gamma=(1.7/\Gamma)$\hmpc\ and 
$\Gamma=\Omega_m h$. The 
normalisation parameter $A$ is fixed through the value of 
$\sigma_8 \equiv \sigma(R=8 h^{-1}{\rm Mpc})$ 
We take the values of 
$\sigma_8$ to be given by \shortcite{ecf96}
\be
\sigma_8 =\left\{ \begin{array}{ll}
              (0.52 \pm 0.04) \Omega_m^{-0.46+0.10 \Omega_m} 
                                &\mbox{(if $\Omega_{\Lambda}=0$)}\\
              (0.52 \pm 0.04) \Omega_m^{-0.52+0.13 \Omega_m}
                                &\mbox{(if $\Omega_{\Lambda}=1-\Omega_m$)}
                \end{array} \right.
\e

Given the above parameters, we can calculate the Press-Schechter 
mass function. However, 
since we prefer to work in terms of $v_c$ rather than $M$, we need to 
know the quantity $\Delta_{\rm vir}(z)$ (see equation (\ref{mass_vel})). For 
flat models with a cosmological constant, this is given by the 
fitting formula \cite{bn98}
\be
\Delta_{\rm vir}(z) = 18 \pi^2 + 82 x - 39 x^2; ~~ 
x \equiv \Omega_m(z)-1
\e

\begin{figure*}
\psfig{figure=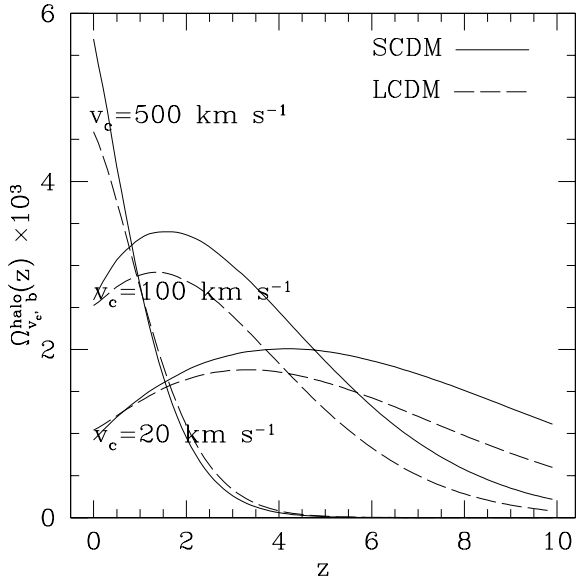,width=8cm}
\caption[]{\label{press_schechter}
The fraction $\Omega_{v_c,b}^{\rm halo}(z)$ for two cosmological models and 
for three values of $v_c$ as indicated in the figure.}
\end{figure*}

In Fig. \ref{press_schechter}, we plot the quantity 
$\Omega_{v_c,b}^{\rm halo}(z)$ for 
the two different cosmological models and for three different values of 
$v_c$. It is clear that haloes of lower circular velocities contain more 
mass at higher redshifts.

There is quite a bit of uncertainty about the SFR. This is mainly because 
of our lack of knowledge of the extinction due to dust.
The extinction has been modelled in various ways (see, for example, 
\citeNP{sagdp99,as00,glp00,hchc01}). In this work, 
we take the 
SFR data obtained from ultraviolet luminosity observations 
\cite{llhc96,csdsb97,mfdgsf96,sagdp99}, corrected for 
extinction by \cite{sagdp99}. The data can be fitted 
with a function of the form 
\be
\dot{\rho}_*(z)=\f{a~ e^{b z}}{e^{c z}+d } 
\e
with the parameters given by 
\be 
a=0.13 ~ M_{\odot} ~ {\rm yr}^{-1} ~ {\rm Mpc}^{-3},~ 
b=2.2,~ c=2.2,~ d=6.0
\label{abcd}
\e

\begin{figure*}
\psfig{figure=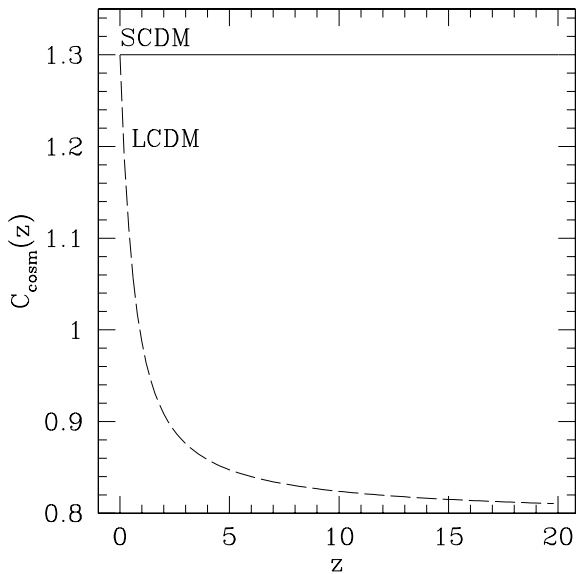,width=8cm}
\caption[]{\label{cosm_correc}
The correction factor for the observed SFR because of using a different 
cosmological model. The observed data points are for a flat $\Omega_m=1, ~ 
h=0.5$ universe, while we are interested in cosmological models 
with different parameters (see text for the cosmological models considered 
in this paper). The correction factor is plotted as a function 
of $z$ for the two cosmological models of our interest (SCDM and LCDM). 
The solid line denotes the correction for SCDM model 
(arising because we use $h=0.65$), while the dashed 
one is for the LCDM model.
}
\end{figure*}

There are two effects which may modify the form and the absolute value 
of this SFR. The first modification is due to cosmology. The above data is 
for a flat $\Omega_m=1$ universe, with $h=0.5$. In order to use the 
same points for a different cosmology, 
one has to note that the SFR is proportional 
to luminosity per volume, which in turn is inversely proportional to the 
distance for a given redshift. 
This means, while considering a flat universe with a cosmological constant 
($\Omega_m + \Omega_{\Lambda} = 1$) 
we have to multiply the above SFR by a correction 
factor
\be
C_{\rm cosm}(z) = \f{h}{0.5} ~\f{2 [1-(1+z)^{-1/2}]}
{\int_0^z \de z' ~ [(1-\Omega_m) + \Omega_m (1+z')^3]^{-1/2}}
\e
In Fig \ref{cosm_correc} we plot this correction factor for the two 
cosmological models of our interest. The correction for 
SCDM model comes 
just because we are using $h=0.65$, and is 
independent of $z$. 

The second effect is due to the stellar initial mass function (IMF). 
The SFR is usually calculated from the observed luminosity using 
some particular IMF. The SFR data we are using is calculated using 
the standard Salpeter IMF with a slope $\Gamma=-1.35$ 
for a mass range $100 M_{\odot}$ -- $0.1M_{\odot}$ 
\cite{mfdgsf96,sagdp99}. However, the real 
IMF  probably
becomes flatter below $1 M_{\odot}$. 
\citeN{leitherer98} calculates that the extrapolation 
of the Salpeter IMF to $0.1M_{\odot}$ overestimates the SFR by a 
factor of 2.5. In this work, we take into account the flattening
and eventual turning over of the IMF below $1 M_{\odot}$ by 
simply reducing the SFR by a factor 2.5.

\begin{figure*}
\psfig{figure=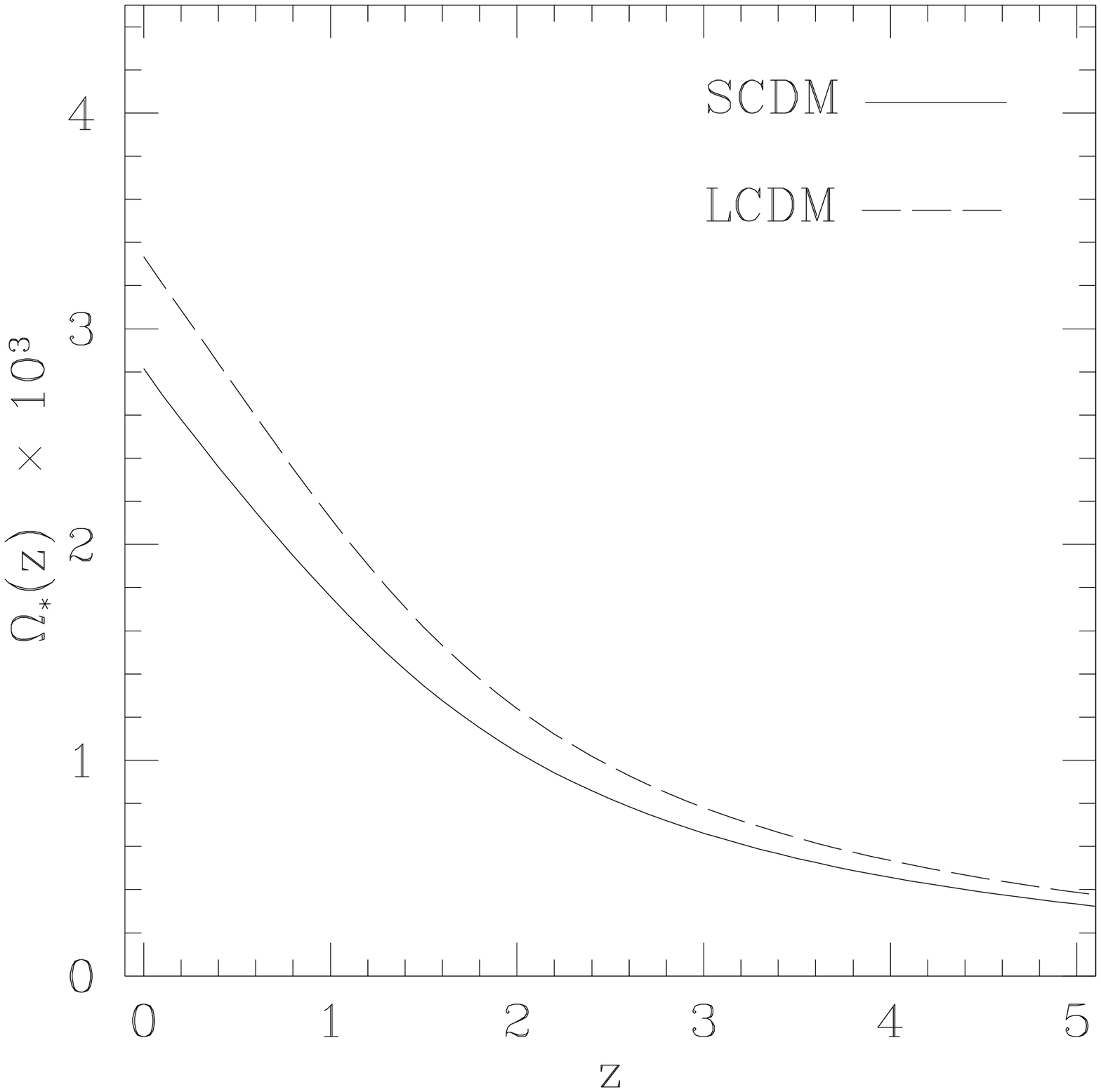,width=8cm}
\caption[]{\label{omega_star}
The density of matter contained in stars $\Omega_*$ as a function of 
$z$, plotted for two cosmological models. 
}
\end{figure*}

Implementing both the corrections, we get the corrected SFR in the form
\be
\dot{\rho}_*(z)=\f{a~ e^{b z}}{e^{c z}+d } ~ \f{C_{\rm cosm}(z)}{2.5}
\e
with the parameters $a,b,c,d$ given in equation (\ref{abcd}).
Given $\dot{\rho}_*(z)$, we can calculate the quantity $\Omega_*(z)$ 
using equations (\ref{rho_star_rhodot}) and (\ref{omega_star_rho_star}). 
This quantity is plotted in 
Fig. \ref{omega_star}, for two cosmological models.

\begin{figure*}
\psfig{figure=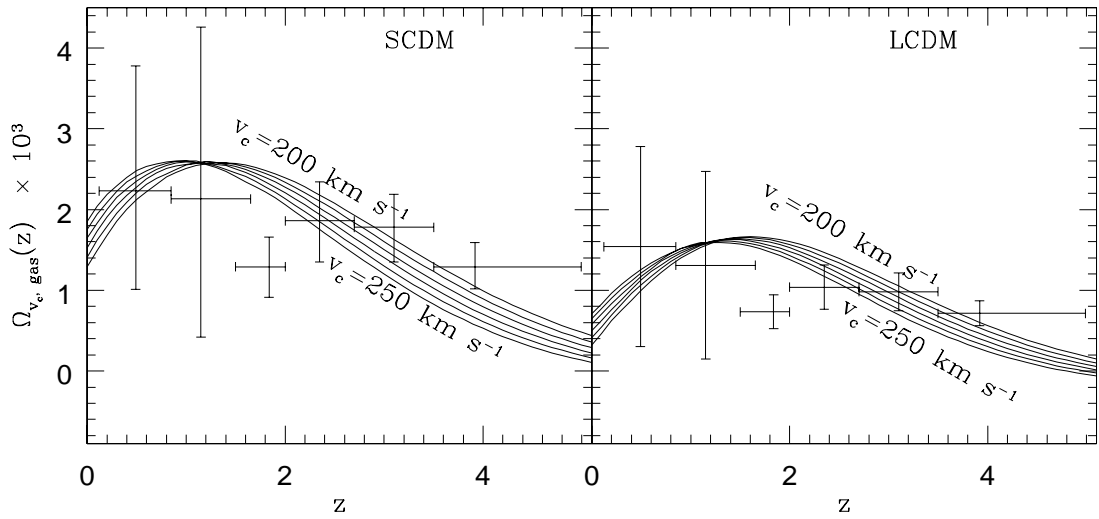,width=16cm}
\caption[]{\label{omega_gas_6vc}
The density of matter contained in gaseous form $\Omega_{v_c, {\rm gas}}(z)$ 
as a function of 
$z$, plotted for two cosmological models. The circular velocity ranges 
from  $v_c=200$ km s$^{-1}$ to $v_c=250$ km s$^{-1}$ in both the plots.  
}
\end{figure*}

We now obtain the main result of our work, 
$\Omega_{v_c,{\rm gas}}(z)$, and compare
it with DLA observations. 
We have already discussed how to compute the 
quantities $\Omega_{v_c,b}^{\rm halo}(z)$ (see equation (\ref{omega_m_b_z})) 
and $\Omega_*$ 
(see equations (\ref{rho_star_rhodot}) and (\ref{omega_star_rho_star})). 
Given these two quantities, 
we can analytically compute  
$\Omega_{v_c,{\rm gas}}(z)$ using 
equation (\ref{omega_m_b_halo_z}).
The observed data used for comparing our model is 
obtained from \citeN{pmsi01}. 
First we consider a scenario in which 
the circular velocities of the collapsed 
haloes do not have significant evolution, i.e., they are more or less 
constant over the redshift range. To be more precise, 
we consider a velocity range 
from $v_c=200$ km s$^{-1}$ to $v_c=250$ km s$^{-1}$ for both the 
cosmological models and see how they compare with the observed points. 
Fig. \ref{omega_gas_6vc} shows 
the comparison within this velocity range.
As can be seen from the figure, the ball park estimate of 
$\Omega_{v_c,{\rm gas}}(z)$ from our simple model is well within 
observational constraints. Furthermore, the general trend of 
evolution of $\Omega_{v_c,{\rm gas}}(z)$ is also 
reproduced.

In passing, we mention that we have plotted the mass density 
$\Omega_{v_c,{\rm gas}}(z)$ contributed by haloes within a 
logarithmic velocity range. We could have as well plotted the quantity 
$\Omega_{\rm gas}(z)$, which is the mass density contributed 
by haloes having a wide range of circular velocities 
(say, from $v_{c,{\rm min}}=50$ km s$^{-1}$ 
to $v_{c,{\rm max}}=250$ km s$^{-1}$) (see \citeN{mm94}). Mathematically, 
this quantity can be obtained from $\Omega_{v_c,{\rm gas}}(z)$ 
through the relation
\be
\Omega_{\rm gas}(z) = \int_{v_{c,{\rm min}}}^{v_{c,{\rm max}}} 
\Omega_{v_c, {\rm gas}}(z) \f{\de \ln M}{\de \ln v_c} ~\de \ln v_c
\e
However, when compared with observations, this quantity 
turns out to be much higher than the observed mass density 
contributed by DLAs. This means that either (i) most of the  
observed DLAs are hosted by DM haloes with a small 
range of circular velocities or, (ii) we are severely underestimating 
the SFR. At this stage, it is quite difficult to distinguish between 
the two possibilities. 

Let us now return back to Fig. \ref{omega_gas_6vc}. 
One notices that observational 
points in higher redshifts are better fitted by curves with lower 
velocities. For example, for the LCDM model, the point at
$z=2.35$ (fourth from left) has a better match with the curve having 
$v_c=250$ km s$^{-1}$ (a curve with a higher $v_c$ will 
do still better), while the point at the highest redshift ($z=3.92$) 
can be fitted with a curve of lower velocity ($v_c < 200$ km s$^{-1}$). 
This might motivate one to consider a second scenario where the 
circular velocity falls as redshift increases.

\begin{figure*}
\psfig{figure=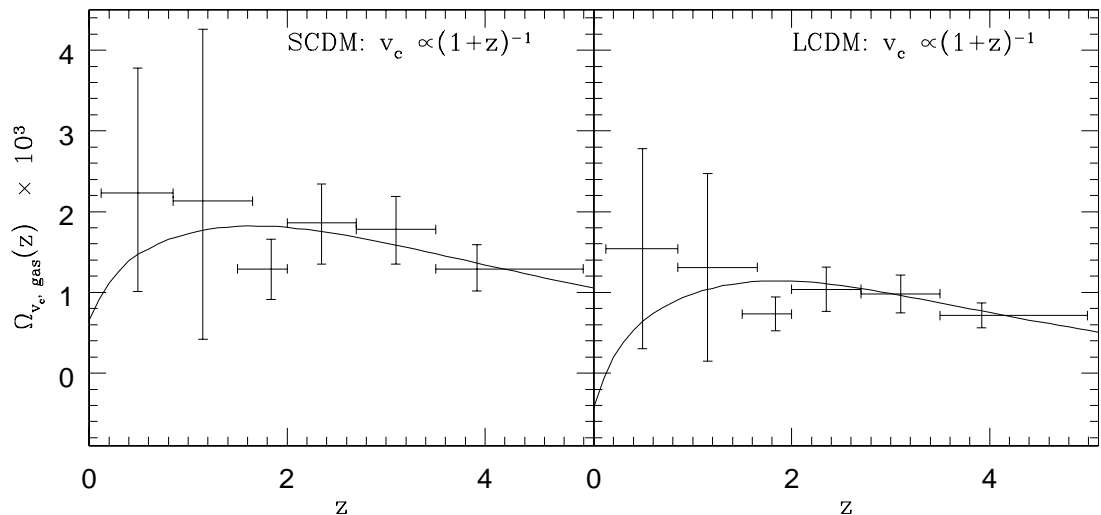,width=16cm}
\caption[]{\label{omega_gas_vc_z}
The density of matter contained in gaseous form $\Omega_{v_c, {\rm gas}}(z)$ 
as a function of 
$z$, plotted for two cosmological models. The circular velocity is 
assumed to be proportional to the scale factor. It is evident from 
the figure that this $z$-dependence 
of $v_c$ does not hold for $z<1$.
}
\end{figure*}

It should be obvious 
that it is impossible to constrain the evolutionary pattern of $v_c$
with such large error bars in the observational data. However, we shall 
illustrate how it can be done in principle. We consider 
a simple evolution of $v_c$ parametrised as
\be
\f{v_c(z)}{250~{\rm km~s^{-1}}} = \eta \left(\f{1+z}{3}\right)^{-1}
\e
where $\eta$ is a parameter of the order unity, to be fixed by 
comparing with observations.
Fig. \ref{omega_gas_vc_z} shows the comparison between our model and 
observations when $v_c$ is given by the above relation. The values 
of $\eta$ for the two cosmological models are
\be
\eta  \simeq \left\{ \begin{array}{ll}
		1.167
	&\mbox{(for SCDM)}\\
		1.163
	&\mbox{(for LCDM)}
                \end{array} \right.
\e
The simple evolution law of $v_c$ matches with observations reasonably 
well, especially at $z>2$. The match is not that good for $z<1$, and 
one has to introduce more complicated redshift dependence to take care 
of this. We do not perform such exercise in this work, as it is 
difficult to constrain any parameter with such large observational errors.

We are thus able to show that our model, in spite of being extremely simple, 
is able to reproduce the broad observational trends for the 
quantity $\Omega_{v_c, {\rm gas}}(z)$ 
(the density of baryons contained within collapsed haloes in gaseous form).
The model can, in principle, be used to predict the circular velocities 
of the collapsed haloes as a function of $z$.

It should be noted that the detailed model of \citeN{kauffmann96} gives 
somewhat
lower circular velocities compared to ours. At this stage 
it is difficult to compare our results directly 
with those of \citeN{kauffmann96}. Since
both have used the PS to calculate the mass in collapsed objects, the
difference must lie in the estimation of SFR. We have used an observed
estimate for the SFR. It is not obvious from the work of \citeNP{kauffmann96} 
how the results for SFR match with observations.

As an extension of this work, 
we also calculate the distribution of circular 
velocities of the DM haloes, using the Press-Schechter formalism 
\cite{hsr98,hsr00}. However, we preferred to keep this paper 
simple and focused on one key idea only.

\section{Discussion}
\label{sec:disc}

We have presented a simple analytical model for estimating amount 
of matter left in gaseous form within the collapsed DM haloes. 
The only ingredients for this model are (i) the Press-Schechter formalism 
for the collapse of DM haloes, and (ii) the observational estimates 
of the SFR at different redshifts. Our model reproduces the mass 
contained in the baryonic gas which has 
cooled within the DM haloes as 
expected in observations. Furthermore, 
the broad trends seen 
in the observed data are reproduced in our model. 
We have used it to estimate the 
circular velocities of the collapsed haloes at different redshifts, which 
might be compared with future observations. 

The model has a couple of important 
uncertainties, described below:

(i) The observational estimate of the SFR is quite uncertain, mainly because 
of the extinction. There are different models for the extinction in the 
literature, and the resulting estimate differ 
quite a bit \cite{sagdp99,hchc01}. 
It is difficult to prove what is 
the best method to use at high redshift for estimating the 
extinction, since there are so
few constraints. In our model 
we have used the extinction 
corrected data of \citeN{sagdp99}, as their estimates 
are consistent with the sub-mm background \cite{as00}. 

Note that this uncertainty may have a large influence on the values of 
$v_c$ calculated in this paper. 
For example, the SFR estimated by \citeN{hchc01} 
is about 5 to 6 times higher than what we have 
used at $z > 2.5$. Using such a SFR would have lead us to use a higher 
value of baryonic matter density so as to match the observations of 
$\Omega_{v_c, {\rm gas}}(z)$. The only obvious way to increase the total 
baryonic matter is to 
include the haloes with lower 
velocities. Thus there is a possibility that the haloes with lower $v_c$ 
than what is calculated in this paper might also 
contribute to $\Omega_{v_c, {\rm gas}}(z)$. Thus, the absolute values 
of $v_c$ predicted in this work 
might have errors. However, the broad features (like $v_c$ decreases with 
increasing $z$) are expected to be still valid.

(ii) In our model, we have neglected the feedback from supernovae and 
stellar winds because there is no simple way to deal with them. 
This might have resulted in a slight underestimate of 
$\Omega_{v_c, {\rm gas}}(z)$. However, we do not expect this correction 
to modify the broad trends of the model. 

We have found that our model matches the observations in the redshift range 
$2 < z < 5$, provided the circular velocities of the 
haloes are within the range 100~km~s$^{-1} \le v_c \le 250$~km~s$^{-1}$.
\citeN{hsr98} and \citeN{hsr00} compare hydrodynamical simulations 
and observed data \cite{pw97} 
to conclude that the typical circular velocities of 
the haloes at high redshifts are $\sim$ 100--200~km~s$^{-1}$. This value is 
in agreement with our analysis. 
Also, kinematic studies of \citeN{lpbws98} 
find that there is a trend for the velocity to decrease with redshift. 
The circular velocity cannot be predicted 
for lower redshifts ($z<2$) because of the large errors on the observed 
$\Omega_{v_c, {\rm gas}}(z)$.

\section*{Acknowledgment}
We gratefully acknowledge the support from the Indo-French Centre for
Promotion of Advanced  Research under contract No. 1710-1. 
We thank Max Pettini and C. Steidel for providing us with the observed data 
used in this work, and for various  useful comments 
on the paper. We also thank K. Subramanian for 
useful comments.
TRC is supported by the University Grants Commission, India.

\bibliography{apjmnemonic,astropap}

\begin{thebibliography}{}

\bibitem[\protect\citeauthoryear{{Adelberger} \& {Steidel}}{{Adelberger} \&
  {Steidel}}{2000}]{as00}
{Adelberger}, K.~L.,  \& {Steidel}, C.~C. 2000, ApJ, 544, 218

\bibitem[\protect\citeauthoryear{{Bi} \& {Davidsen}}{{Bi} \&
  {Davidsen}}{1997}]{bd97}
{Bi}, H.,  \& {Davidsen}, A.~F. 1997, ApJ, 479, 523

\bibitem[\protect\citeauthoryear{{Bryan} \& {Norman}}{{Bryan} \&
  {Norman}}{1998}]{bn98}
{Bryan}, G.~L.,  \& {Norman}, M.~L. 1998, ApJ, 495, 80

\bibitem[\protect\citeauthoryear{{Choudhury}, {Padmanabhan}, \&
  {Srianand}}{{Choudhury} et~al.}{2001}]{cps01}
{Choudhury}, T.~R., {Padmanabhan}, T.,  \& {Srianand}, R. 2001, MNRAS, 322, 561

\bibitem[\protect\citeauthoryear{{Choudhury}, {Srianand}, \&
  {Padmanabhan}}{{Choudhury} et~al.}{2001}]{csp01}
{Choudhury}, T.~R., {Srianand}, R.,  \& {Padmanabhan}, T. 2001, ApJ, 559, 29

\bibitem[\protect\citeauthoryear{{Connolly} et~al.}{{Connolly}
  et~al.}{1997}]{csdsb97}
{Connolly}, A.~J., {Szalay}, A.~S., {Dickinson}, M., {Subbarao}, M.~U.,  \&
  {Brunner}, R.~J. 1997, ApJ, 486, L11

\bibitem[\protect\citeauthoryear{{Efstathiou}}{{Efstathiou}}{2000}]{efstathiou%
00}
{Efstathiou}, G. 2000, MNRAS, 317, 697

\bibitem[\protect\citeauthoryear{{Efstathiou}, {Bond}, \& {White}}{{Efstathiou}
  et~al.}{1992}]{ebw92}
{Efstathiou}, G., {Bond}, J.~R.,  \& {White}, S.~D.~M. 1992, MNRAS, 258, 1P

\bibitem[\protect\citeauthoryear{{Eke}, {Cole}, \& {Frenk}}{{Eke}
  et~al.}{1996}]{ecf96}
{Eke}, V.~R., {Cole}, S.,  \& {Frenk}, C.~S. 1996, MNRAS, 282, 263

\bibitem[\protect\citeauthoryear{{Gardner} et~al.}{{Gardner}
  et~al.}{1997a}]{gkhw97}
{Gardner}, J.~P., {Katz}, N., {Hernquist}, L.,  \& {Weinberg}, D.~H. 1997a,
  ApJ, 484, 31

\bibitem[\protect\citeauthoryear{{Gardner} et~al.}{{Gardner}
  et~al.}{1997b}]{gkwh97}
{Gardner}, J.~P., {Katz}, N., {Weinberg}, D.~H.,  \& {Hernquist}, L. 1997b,
  ApJ, 486, 42

\bibitem[\protect\citeauthoryear{{Gispert}, {Lagache}, \& {Puget}}{{Gispert}
  et~al.}{2000}]{glp00}
{Gispert}, R., {Lagache}, G.,  \& {Puget}, J.~L. 2000, A\&A, 360, 1

\bibitem[\protect\citeauthoryear{{Haehnelt}, {Steinmetz}, \&
  {Rauch}}{{Haehnelt} et~al.}{1998}]{hsr98}
{Haehnelt}, M.~G., {Steinmetz}, M.,  \& {Rauch}, M. 1998, ApJ, 495, 647

\bibitem[\protect\citeauthoryear{{Haehnelt}, {Steinmetz}, \&
  {Rauch}}{{Haehnelt} et~al.}{2000}]{hsr00}
{Haehnelt}, M.~G., {Steinmetz}, M.,  \& {Rauch}, M. 2000, ApJ, 534, 594

\bibitem[\protect\citeauthoryear{{Hopkins} et~al.}{{Hopkins}
  et~al.}{2001}]{hchc01}
{Hopkins}, A.~M., {Connolly}, A.~J., {Haarsma}, D.~B.,  \& {Cram}, L.~E. 2001,
  AJ, 122, 288

\bibitem[\protect\citeauthoryear{{Jedamzik} \& {Prochaska}}{{Jedamzik} \&
  {Prochaska}}{1998}]{jp98}
{Jedamzik}, K.,  \& {Prochaska}, J.~X. 1998, MNRAS, 296, 430

\bibitem[\protect\citeauthoryear{{Katz} et~al.}{{Katz} et~al.}{1996}]{kwhm96}
{Katz}, N., {Weinberg}, D.~H., {Hernquist}, L.,  \& {Miralda-Escude}, J. 1996,
  ApJ, 457, L57

\bibitem[\protect\citeauthoryear{{Kauffmann}}{{Kauffmann}}{1996}]{kauffmann96}
{Kauffmann}, G. 1996, MNRAS, 281, 475

\bibitem[\protect\citeauthoryear{{Kauffmann} \& {Charlot}}{{Kauffmann} \&
  {Charlot}}{1994}]{kc94}
{Kauffmann}, G.,  \& {Charlot}, S. 1994, ApJ, 430, L97

\bibitem[\protect\citeauthoryear{{Lanzetta}, {Wolfe}, \& {Turnshek}}{{Lanzetta}
  et~al.}{1995}]{lwt95}
{Lanzetta}, K.~M., {Wolfe}, A.~M.,  \& {Turnshek}, D.~A. 1995, ApJ, 440, 435

\bibitem[\protect\citeauthoryear{{Ledoux} et~al.}{{Ledoux}
  et~al.}{1998}]{lpbws98}
{Ledoux}, C., {Petitjean}, P., {Bergeron}, J., {Wampler}, E.~J.,  \&
  {Srianand}, R. 1998, A\&A, 337, 51

\bibitem[\protect\citeauthoryear{{Leitherer}}{{Leitherer}}{1998}]{leitherer98}
{Leitherer}, C. 1998, STScI preprint No. 1254

\bibitem[\protect\citeauthoryear{{Lilly} et~al.}{{Lilly} et~al.}{1996}]{llhc96}
{Lilly}, S.~J., {Le Fevre}, O., {Hammer}, F.,  \& {Crampton}, D. 1996, ApJ,
  460, L1

\bibitem[\protect\citeauthoryear{{Ma} \& {Bertschinger}}{{Ma} \&
  {Bertschinger}}{1994}]{mb94}
{Ma}, C.,  \& {Bertschinger}, E. 1994, ApJ, 434, L5

\bibitem[\protect\citeauthoryear{{Ma} et~al.}{{Ma} et~al.}{1997}]{mbhwk97}
{Ma}, C., {Bertschinger}, E., {Hernquist}, L., {Weinberg}, D.~H.,  \& {Katz},
  N. 1997, ApJ, 484, L1

\bibitem[\protect\citeauthoryear{{Madau} et~al.}{{Madau}
  et~al.}{1996}]{mfdgsf96}
{Madau}, P., {Ferguson}, H.~C., {Dickinson}, M.~E., {Giavalisco}, M.,
  {Steidel}, C.~C.,  \& {Fruchter}, A. 1996, MNRAS, 283, 1388

\bibitem[\protect\citeauthoryear{{Maller} et~al.}{{Maller}
  et~al.}{2001}]{mpsp01}
{Maller}, A.~H., {Prochaska}, J.~X., {Somerville}, R.~S.,  \& {Primack}, J.~R.
  2001, MNRAS, 326, 1475

\bibitem[\protect\citeauthoryear{{Miralda-Escude} et~al.}{{Miralda-Escude}
  et~al.}{1996}]{mcor96}
{Miralda-Escude}, J., {Cen}, R., {Ostriker}, J.~P.,  \& {Rauch}, M. 1996, ApJ,
  471, 582

\bibitem[\protect\citeauthoryear{{Mo} \& {Miralda-Escude}}{{Mo} \&
  {Miralda-Escude}}{1994}]{mm94}
{Mo}, H.~J.,  \& {Miralda-Escude}, J. 1994, ApJ, 430, L25

\bibitem[\protect\citeauthoryear{{Peroux} et~al.}{{Peroux}
  et~al.}{2001}]{pmsi01}
{Peroux}, C., {McMahon}, R.~G., {Storrie-Lombardi}, L.~J.,  \& {Irwin}, M.~J.
  2001, Preprint: astro-ph/0107045

\bibitem[\protect\citeauthoryear{{Pettini} et~al.}{{Pettini}
  et~al.}{1999}]{pesb99}
{Pettini}, M., {Ellison}, S.~L., {Steidel}, C.~C.,  \& {Bowen}, D.~V. 1999,
  ApJ, 510, 576

\bibitem[\protect\citeauthoryear{{Pettini} et~al.}{{Pettini}
  et~al.}{1997a}]{pksh97}
{Pettini}, M., {King}, D.~L., {Smith}, L.~J.,  \& {Hunstead}, R.~W. 1997a, ApJ,
  478, 536

\bibitem[\protect\citeauthoryear{{Pettini} et~al.}{{Pettini}
  et~al.}{1994}]{pshk94}
{Pettini}, M., {Smith}, L.~J., {Hunstead}, R.~W.,  \& {King}, D.~L. 1994, ApJ,
  426, 79

\bibitem[\protect\citeauthoryear{{Pettini} et~al.}{{Pettini}
  et~al.}{1997b}]{pskh97}
{Pettini}, M., {Smith}, L.~J., {King}, D.~L.,  \& {Hunstead}, R.~W. 1997b, ApJ,
  486, 665

\bibitem[\protect\citeauthoryear{{Press} \& {Schechter}}{{Press} \&
  {Schechter}}{1974}]{ps74}
{Press}, W.~H.,  \& {Schechter}, P. 1974, MNRAS, 187, 425

\bibitem[\protect\citeauthoryear{{Prochaska} \& {Wolfe}}{{Prochaska} \&
  {Wolfe}}{1997}]{pw97}
{Prochaska}, J.~X.,  \& {Wolfe}, A.~M. 1997, ApJ, 487, 73

\bibitem[\protect\citeauthoryear{{Prochaska} \& {Wolfe}}{{Prochaska} \&
  {Wolfe}}{1998}]{pw98}
{Prochaska}, J.~X.,  \& {Wolfe}, A.~M. 1998, ApJ, 507, 113

\bibitem[\protect\citeauthoryear{{Somerville} \& {Primack}}{{Somerville} \&
  {Primack}}{1999}]{sp99}
{Somerville}, R.~S.,  \& {Primack}, J.~R. 1999, MNRAS, 310, 1087

\bibitem[\protect\citeauthoryear{{Steidel} et~al.}{{Steidel}
  et~al.}{1999}]{sagdp99}
{Steidel}, C.~C., {Adelberger}, K.~L., {Giavalisco}, M., {Dickinson}, M.,  \&
  {Pettini}, M. 1999, ApJ, 519, 1

\bibitem[\protect\citeauthoryear{{Storrie-Lombardi}, {McMahon}, \&
  {Irwin}}{{Storrie-Lombardi} et~al.}{1996}]{smi96}
{Storrie-Lombardi}, L.~J., {McMahon}, R.~G.,  \& {Irwin}, M.~J. 1996, MNRAS,
  283, L79

\bibitem[\protect\citeauthoryear{{Subramanian} \& {Padmanabhan}}{{Subramanian}
  \& {Padmanabhan}}{1994}]{sp94}
{Subramanian}, K.,  \& {Padmanabhan}, T. 1994, Preprint: astro-ph/9402006

\bibitem[\protect\citeauthoryear{{White} \& {Frenk}}{{White} \&
  {Frenk}}{1991}]{wf91}
{White}, S.~D.~M.,  \& {Frenk}, C.~S. 1991, ApJ, 379, 52

\bibitem[\protect\citeauthoryear{{White} \& {Rees}}{{White} \&
  {Rees}}{1978}]{wr78}
{White}, S.~D.~M.,  \& {Rees}, M.~J. 1978, MNRAS, 183, 341

\bibitem[\protect\citeauthoryear{{Wolfe} et~al.}{{Wolfe} et~al.}{1986}]{wtsc86}
{Wolfe}, A.~M., {Turnshek}, D.~A., {Smith}, H.~E.,  \& {Cohen}, R.~D. 1986,
  ApJS, 61, 249

\end{thebibliography}

\bibliographystyle{apj}

\label{lastpage}

\end{document}